\documentclass[final,5p,times]{elsarticle}
\usepackage{amssymb}
\usepackage{lineno}

\usepackage{graphicx}
\usepackage{comment}
\usepackage{here}
\usepackage{subcaption}
\usepackage{amsmath}
\usepackage{adjustbox}

\journal{Nuclear Instruments and Methods in Physics Research Section A}

\begin{document}

\begin{frontmatter}

%% Title, authors and addresses

%% use the tnoteref command within \title for footnotes;
%% use the tnotetext command for theassociated footnote;
%% use the fnref command within \author or \address for footnotes;
%% use the fntext command for theassociated footnote;
%% use the corref command within \author for corresponding author footnotes;
%% use the cortext command for theassociated footnote;
%% use the ead command for the email address,
%% and the form \ead[url] for the home page:
%% \title{Title\tnoteref{label1}}
%% \tnotetext[label1]{}
%% \author{Name\corref{cor1}\fnref{label2}}
%% \ead{email address}
%% \ead[url]{home page}
%% \fntext[label2]{}
%% \cortext[cor1]{}
%% \affiliation{organization={},
%%             addressline={},
%%             city={},
%%             postcode={},
%%             state={},
%%             country={}}
%% \fntext[label3]{}

\title{Irradiation Studies of TGC Electronics Components for the ATLAS Experiment at High-Luminosity LHC}

\author[Nagoya1]{Yuya Ohsumi}
\author[Nagoya1]{Daisuke Hashimoto}

\author[Nagoya1,Nagoya2]{Yasuyuki Horii\corref{correspondingauthor}}
\cortext[correspondingauthor]{Corresponding author}
\ead{yhorii@hepl.phys.nagoya-u.ac.jp}

\author[Tokyo]{Takumi Aoki}
\author[Nagoya1]{Haruka Asada}
\author[Nagoya1]{Kazumasa Hashizume}
\author[Nagoya1]{Hayato Inaguma}
\author[ICEPP]{Masaya Ishino}
%\author[Kobe1]{Masato Kanasaki}
\author[Nagoya1]{Miyuki Kikuchi}
\author[Tokyo]{Shota Kondo}
\author[Tokyo]{Reita Maeno}
\author[Tokyo]{Airu Makita}
\author[Nagoya1]{Masaki Minakawa}
\author[Nagoya1]{Yuki Mitsumori}
\author[Nagoya1]{Yuki Nabeyama}
\author[Tokyo]{Ren Nagasaka}
\author[Nagoya1]{Takumi Nakajima}
\author[Tokyo]{Yoshifumi Narukawa}
\author[Kobe2]{Atsuhiko Ochi}
\author[ICEPP]{Yasuyuki Okumura}
\author[KEK]{Osamu Sasaki}
\author[Tokyo]{Aoto Tanaka}
\author[Kobe1]{Akira Taniike}
\author[Nagoya1,Nagoya2,KEK]{Makoto Tomoto}
\author[Nagoya1]{Arisa Wada}
\author[Tokyo]{and Erika Yamashita}

%\author[nagoya]{Yuya Ohsumi \corref{nagoya}, Yasuyuki Horii, Atsuhiko Ochi, Aoto Tanaka, Arisa Wada, Daisuke Hashimoto, Erika Yamashita, Haruka Asada, Hayato Inaguma, Kazumasa Hashizume, Makoto Tomoto, Masaki Minakawa, Masaya Ishino, Miyuki Kikuchi, Osamu Sasaki, Ren Nagasaka, Takumi Aoki, Takumi Nakajima, Yasuyuki Okumura, Toshifumi Narukawa, Yuki Mitsumori, Yuki Nabeyama}

%% use optional labels to link authors explicitly to addresses:
%% \author[label1,label2]{}

\affiliation[Nagoya1]{
             organisation={Graduate School of Science, Nagoya University},
             addressline={Furo-cho, Chikusa-ku},
             city={Nagoya},
             postcode={464-8602},
             %%state={Aichi},
             country={Japan}}

\affiliation[Nagoya2]{
             organisation={Kobayashi-Maskawa Institute, Nagoya University},
             addressline={Furo-cho, Chikusa-ku},
             city={Nagoya},
             postcode={464-8602},
             %%state={Aichi},
             country={Japan}}

\affiliation[Tokyo]{
             organisation={Department of Physics, The University of Tokyo},
             addressline={Hongo, Bunkyo-ku},
             city={Tokyo},
             postcode={113-0033},
             country={Japan}}

\affiliation[ICEPP]{
             organisation={International Center for Elementary Particle Physics, The University of Tokyo},
             addressline={Hongo, Bunkyo-ku},
             city={Tokyo},
             postcode={113-0033},
             country={Japan}}

\affiliation[Kobe1]{
             organisation={Graduate School of Maritime Sciences, Kobe University},
             addressline={Fukaeminami-machi, Higashinada-ku},
             city={Kobe},
             postcode={658-0022},
             country={Japan}}

\affiliation[Kobe2]{
             organisation={Graduate School of Science, Kobe University},
             addressline={Rokkodai-cho, Nada-ku},
             city={Kobe},
             postcode={657-8501},
             country={Japan}}

\affiliation[KEK]{
             organisation={High Energy Accelerator Research Organization},
             addressline={Oho},
             city={Tsukuba},
             postcode={305-0801},
             country={Japan}}

\begin{abstract}
This paper evaluates the radiation tolerance of commercial off-the-shelf (COTS) electronics components for use in the Thin Gap Chamber (TGC) frontend electronics of the ATLAS experiment at the High-Luminosity LHC (HL-LHC). The ATLAS experiment has accumulated more than 450 $\rm{fb}^{-1}$ of data as of 2025. Its luminosity upgrade, the HL-LHC scheduled to begin operation in 2030, will deliver 3000--4000~$\rm{fb}^{-1}$ over ten years and lead to substantially higher radiation levels in detector electronics. The radiation levels for the TGC frontend electronics are estimated to be 4.1--7.3~Gy in terms of Total Ionizing Dose (TID) and 1.1--$2.2\times 10^{11}$~n$_{\rm1MeV}$~$\rm{cm}^{-2}$ in terms of Non-Ionizing Energy Loss (NIEL). To evaluate component suitability under these conditions, TID tests were conducted using Cobalt-60 gamma rays at Nagoya University, and NIEL tests were performed with the Tandem Accelerator at Kobe University. Various COTS components, including SFP+ optical transceivers, clock jitter cleaners, optical fibers, voltage references, operational amplifiers, analog-to-digital converters, digital-to-analog converters, SD cards, flash memories, and low-dropout regulators, were tested and evaluated against the required radiation levels. The results demonstrate that all evaluated components meet the TID and NIEL tolerance requirements for application in the TGC frontend electronics at the HL-LHC.
%\end{linenumbers}
\end{abstract}

%%Graphical abstract
%\begin{graphicalabstract}
%\includegraphics{grabs}
%\end{graphicalabstract}

%%Research highlights
%\begin{highlights}
%\item Research highlight 1
%\item Research highlight 2
%\end{highlights}

%\begin{keyword}
%% keywords here, in the form: keyword \sep keyword

%% PACS codes here, in the form: \PACS code \sep code

%% MSC codes here, in the form: \MSC code \sep code
%% or \MSC[2008] code \sep code (2000 is the default)

%\end{keyword}

\end{frontmatter}

%\linenumbers

%% main text
\section{Introduction}
The ATLAS experiment~\cite{ATLAS} at the Large Hadron Collider (LHC) aims to investigate the mechanism of electroweak symmetry breaking and to search for physics beyond the Standard Model. The observation of the Higgs boson was reported in 2012~\cite{Higgs}, and its coupling constants to various particles have subsequently been measured~\cite{Coupling}.
A major luminosity upgrade of the LHC, the High-Luminosity Large Hadron Collider (HL-LHC)~\cite{HL-LHC}, is currently under preparation, with operation scheduled to begin in 2030.
The HL-LHC is expected to deliver an integrated luminosity of 3000--4000~$\rm{fb}^{-1}$ over ten years.
%The HL-LHC is expected to achieve a peak luminosity of $5$--$7.5 \times 10^{34} ~ {\rm cm}^{-2} {\rm s}^{-1}$, which is 5--7.5 times higher than the original LHC design value.
The electronics of the muon trigger detector, the Thin Gap Chamber (TGC)~\cite{MUON_TDR}, will be upgraded~\cite{PhaseII_TDR}. The frontend electronics are installed in close proximity to the detector and are therefore required to exhibit high radiation tolerance.

In this work, we evaluated the tolerance of commercial off-the-shelf (COTS) components for the TGC frontend electronics against Total Ionizing Dose (TID) and Non-Ionizing Energy Loss (NIEL). The tested COTS components include SFP+ optical transceivers, clock jitter cleaners, optical fibers, voltage references, operational amplifiers, analog-to-digital converters (ADCs), digital-to-analog converters (DACs), SD cards, flash memories, and low-dropout regulators.

\section{Radiation level of frontend electronics for TGC}
The TGC electronics for the ATLAS experiment at the HL-LHC consist of PS boards and JATHub boards installed in the ATLAS detector area~\cite{PhaseII_TDR}. The radiation levels at their locations have been estimated using a Monte Carlo simulation based on Geant4~\cite{Geant4}.
The PS and JATHub boards are positioned approximately 8 m and 12 m from the beam axis, respectively. Both are located at a longitudinal distance of approximately 11 m from the interaction point along the beam direction.
The simulated radiation level (SRL) is estimated to be of the order of O(1) Gy for TID and O($10^{11}$)~n$_{\rm1MeV}$~$\rm{cm}^{-2}$ for NIEL.
The radiation tolerance criteria (RTC) are defined by applying safety factors (SFs) to the SRL in order to account for various sources of uncertainty. The RTC is calculated as
\begin{equation}
{\rm RTC} = {\rm SF_{sim}} \times {\rm SF_{test}} \times {\rm SF_{lot}},
\end{equation}
where
\begin{itemize}
\item ${\rm SF_{sim}}$ accounts for uncertainties in the Geant4 simulation,
\item ${\rm SF_{test}}$ accounts for differences between the irradiation test environment and the actual experimental conditions,
\item ${\rm SF_{lot}}$ accounts for variations within and between production lots of the chips.
\end{itemize}
Table~\ref{SF_Table} summarizes the values of the SFs for each radiation effect. The resulting RTC for TID and NIEL are summarized in Tables~\ref{RTC_TID} and~\ref{RTC_NIEL}, respectively.

\begin{table}[htbp]
\centering
\caption{Summary of the SFs for each radiation effect. The values given in parentheses correspond to the cases in which chips from the final production lot or reels were tested.}
\begin{adjustbox}{center}
\scalebox{1.0}{
\begin{tabular}{ccccccccc} \hline 
\begin{tabular}[c]{@{}c@{}}$\text{SF}_\text{sim}$\\ TID\end{tabular} & \begin{tabular}[c]{@{}c@{}}$\text{SF}_\text{sim}$\\ NIEL\end{tabular} &  \begin{tabular}[c]{@{}c@{}}$\text{SF}_\text{test}$\\ TID\end{tabular} & \begin{tabular}[c]{@{}c@{}}$\text{SF}_\text{test}$\\ NIEL\end{tabular} & \begin{tabular}[c]{@{}c@{}}$\text{SF}_\text{lot}$\\ TID\end{tabular} & \begin{tabular}[c]{@{}c@{}}$\text{SF}_\text{lot}$\\ NIEL\end{tabular} & \\ \hline \hline
1.5 & 1.5 & 1 & 1 or 1.3 & 3 (1) & 3 (1) \\ \hline
\end{tabular}
}
\end{adjustbox}
\label{SF_Table}
\end{table}

\begin{table}[htbp]
\centering
\caption{Summary of RTC, SRL, and SF for TID. The units of both RTC and SRL are Gray. Values given in parentheses correspond to the case of $\rm{SF}_{\rm{lot}} = 1$. The values in the first and second rows represent those for the PS and JATHub boards, respectively.}
\begin{tabular}{lcccl} \hline 
RTC & SRL & $\text{SF}_\text{sim}$ & $\text{SF}_\text{test}$ & $\text{SF}_\text{lot}$ \\ \hline \hline
18 (6) & 4.1 & 1.5   & 1      & 3 (1) \\ \hline
33 (11) & 7.3 & 1.5   & 1      & 3 (1) \\ \hline
\end{tabular}
\label{RTC_TID}
\end{table}

\begin{table}[htbp]
\centering
\caption{Summary of RTC, SRL, and SF for NIEL. The units of both RTC and SRL are $10^{11}$ n$_{\rm1MeV}$ $\rm{cm}^{-2}$. Values given in parentheses correspond to the case of $\rm{SF}_{\rm{lot}} = 1$. The values in the first and second rows represent those for the PS and JATHub boards, respectively.}
\begin{tabular}{lcccl} \hline 
RTC & SRL & $\text{SF}_\text{sim}$ & $\text{SF}_\text{test}$ & $\text{SF}_\text{lot}$ \\ \hline \hline
7 (2) & 1.1 & 1.5   & 1.3    & 3 (1) \\ \hline
13 (4) & 2.2 & 1.5   & 1.3    & 3 (1) \\ \hline
\end{tabular}
\label{RTC_NIEL}
\end{table}

\section{Targets and test method}

\begin{comment}
\begin{figure}[h]
\begin{center}
     \begin{tabular}{cc}
     \begin{minipage}[b]{0.45\hsize}
     \begin{center}
     \includegraphics[keepaspectratio, scale=0.21]{VrefBoard_zoom.jpg}
     \caption{}
     \label{}
     \end{center}
     \end{minipage}
     
     \begin{minipage}[b]{0.45\hsize}
     \begin{center}
     \includegraphics[keepaspectratio, scale=0.40]{OPamp_board.png}
     \caption{}
     \label{}
     \end{center}
     \end{minipage}
     \end{tabular}
\end{center}
\end{figure}
\end{comment} 

\subsection{TID tests}

We evaluated the tolerance to TID of COTS electronic components that were considered for the TGC frontend electronics. The tested components include SFP+ optical transceivers, clock jitter cleaners, optical fibers, voltage references, operational amplifiers, ADCs, DACs, SD cards, flash memories, and low-dropout regulators.

\begin{figure}[htbp]
    \centering
    \includegraphics[keepaspectratio, scale=0.35]{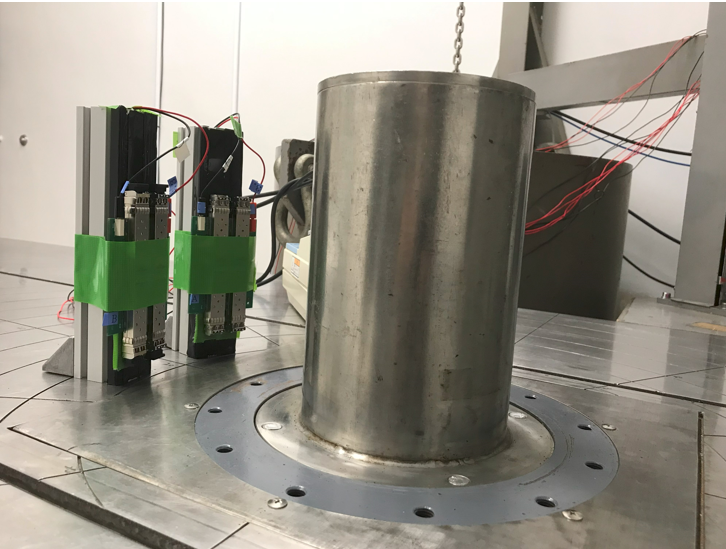}
    \caption{Setup of the TID irradiation tests at Nagoya University. The Cobalt-60 source is housed inside the cylindrical irradiation chamber. Bias voltages are supplied from power supply modules located outside the irradiation area.}
    \label{TIDSetup}
\end{figure}

\begin{figure}[htbp]
\begin{center}
     \begin{tabular}{cc}
     \begin{minipage}[b]{0.45\hsize}
     \begin{center}
     \includegraphics[keepaspectratio, scale=0.50]{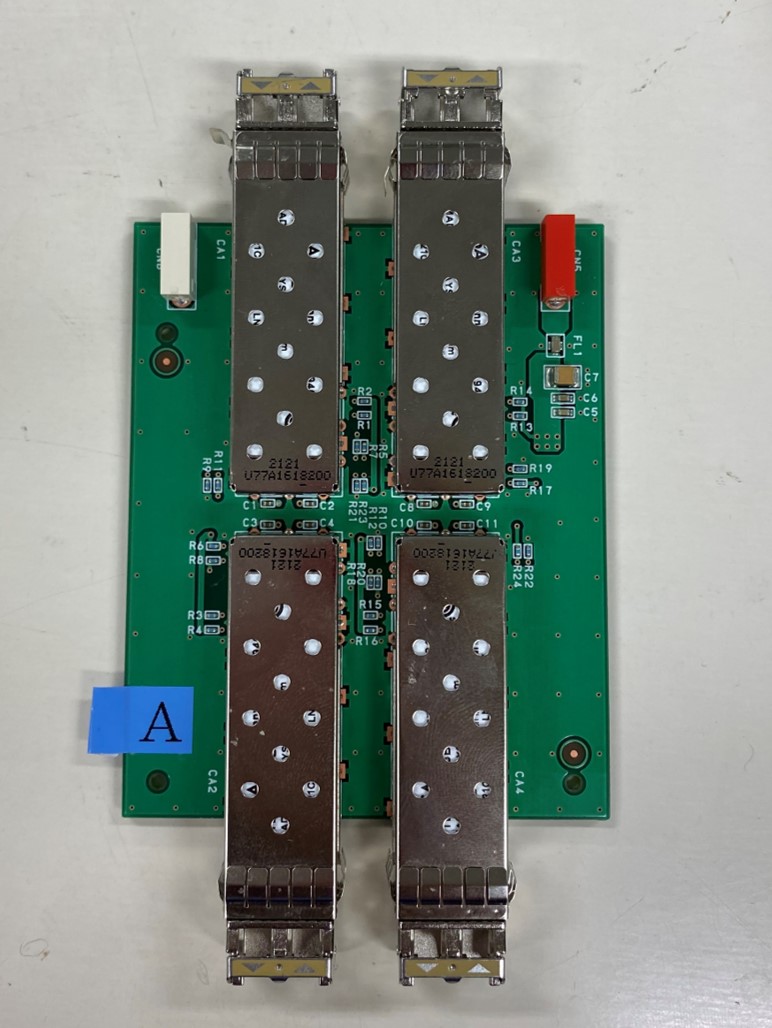}
     \subcaption{}
     \label{SFP+EVB}
     \end{center}
     \end{minipage}
     
     \begin{minipage}[b]{0.45\hsize}
     \begin{center}
     \includegraphics[keepaspectratio, scale=0.38]{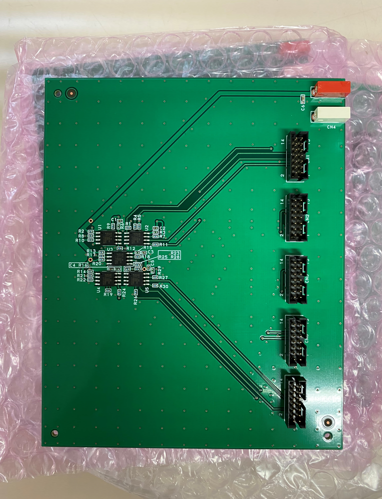}
     \subcaption{}
     \label{QSPIEVM}
     \end{center}
     \end{minipage}
     \end{tabular}
     \caption{(a) Power supply board for SFP+ modules and (b) power supply and evaluation board for flash memory.}
\end{center}
\end{figure}

Gamma-ray irradiation was carried out at the Cobalt-60 facility of Nagoya University (Fig.~\ref{TIDSetup}). To properly account for charge accumulation effects induced by TID, all devices were powered during irradiation.
For the flash memories and SFP+ optical transceivers, dedicated boards were developed to provide stable power during irradiation. Figure~\ref{SFP+EVB} shows the custom power supply board for the SFP+ modules, with four SFP+ devices mounted on each board. Figure~\ref{QSPIEVM} shows the power supply and verification board developed for the flash memory. Using this board, the flash memory was operated and monitored via the SPI protocol during testing.
For the remaining components, manufacturer-provided evaluation boards were used both to supply power during irradiation and to perform functional verification after irradiation. All irradiation and subsequent performance tests were conducted at room temperature.

\subsection{NIEL tests}

\begin{figure}[htbp]
    \centering
    \includegraphics[keepaspectratio, scale=0.30]{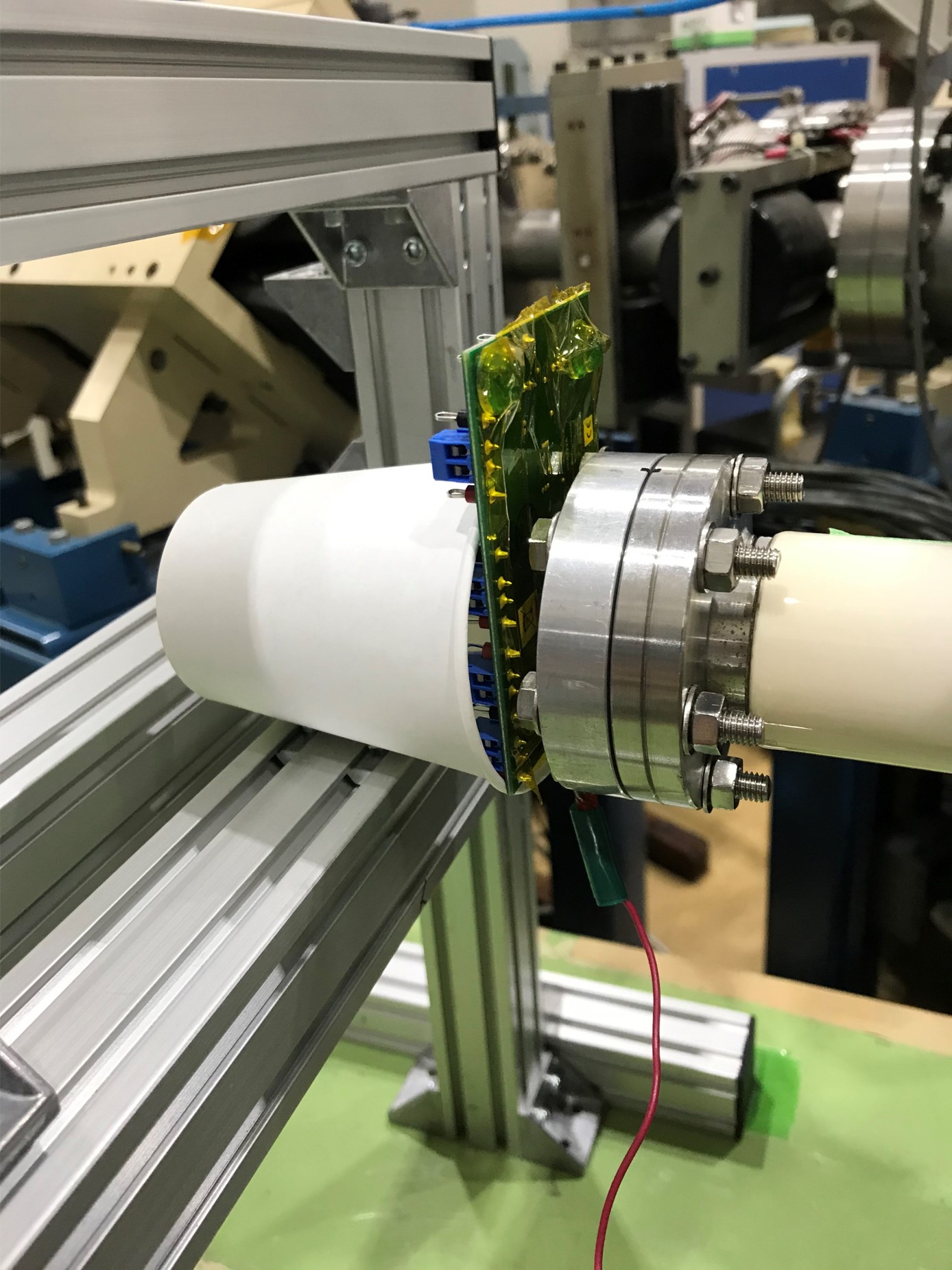}
    \caption{Setup of the NIEL tests near the beam spot. Neutrons are generated by deuterium-ion bombardment of a beryllium target.}
    \label{NIELSetup}
\end{figure}

We evaluated the NIEL tolerance of selected COTS components, SFP+ optical transceivers, voltage references, operational amplifiers, DACs, flash memories, and low-dropout regulators. Neutron irradiation was carried out using the Tandem Accelerator at Kobe University (Fig.~\ref{NIELSetup}). Neutrons were produced via a beryllium (Be) target bombarded by deuterium ions, resulting in a neutron energy spectrum with a peak around 2 MeV~\cite{TwoMeV}. The neutron flux was estimated from the integrated current of the Be target, assuming $(4.9 \pm 1.5)~{\rm MHz/cm}^{2}/\mu{\rm A}$ at a distance of 10 cm from the Be target along the beam axis~\cite{NeutronFlux}. Unlike the TID tests, in which the devices were powered during irradiation to account for charge accumulation effects, the COTS devices were left unpowered during the NIEL tests to account for displacement damage effects. All irradiation and subsequent performance tests were conducted at room temperature.

\begin{comment}
    \begin{figure}[h]
\begin{center}
     \begin{tabular}{cc}
     \begin{minipage}[b]{0.45\hsize}
     \begin{center}
     \includegraphics[keepaspectratio, scale=0.40]{SFP_board.jpg}
     \caption{}
     \label{}
     \end{center}
     \end{minipage}
     
     \begin{minipage}[b]{0.45\hsize}
     \begin{center}
     \includegraphics[keepaspectratio, scale=0.35]{LDO_board.jpg}
     \caption{}
     \label{}
     \end{center}
     \end{minipage}
     \end{tabular}
\end{center}
\end{figure}
\end{comment}

\section{Result of the TID tests}
This section presents the results of the TID tests. The outcomes for individual components are described first, followed by a summary table.

\subsection{SFP+ optical transceivers}

The SFP+ modules irradiated in this study were the AFBR-709SMZ, FTLX8574D3BCV, and FSPP-H7-M85-X3DM. The output light intensity of the transmitters was measured, and communication tests were performed at 8.0~Gbps, the assumed data transfer rate for the TGC electronics, using the Kintex-7 FPGA evaluation kit KC705. The requirement was that the light intensity remain within the receiver acceptance range and that the bit error rate remain negligible when transferring a 31-bit pseudorandom binary sequence. Power cycling was performed at each dose level to evaluate the functionality of the device configuration.

\begin{figure}[htbp]
\begin{center}
\includegraphics[keepaspectratio, scale=0.35]{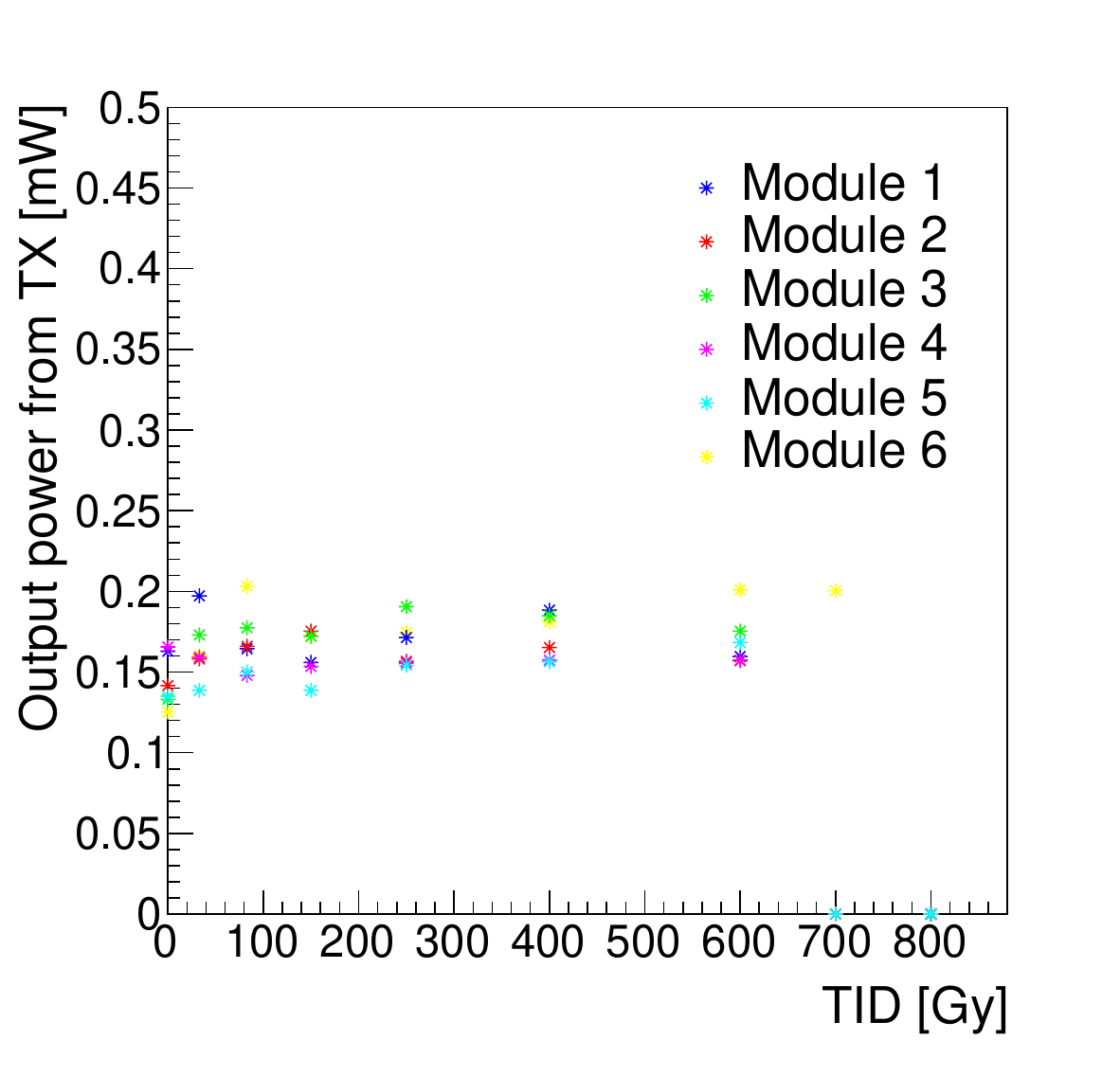}
\caption{Output light intensity of AFBR-709SMZ transmitters during the second test campaign. The output power dropped to 0~mW for five (six) modules after 700~Gy (800~Gy) irradiation.}
\label{SFP_plot_TID}
\end{center}
\end{figure}

A total of eleven AFBR-709SMZ modules were evaluated in two test campaigns. In the first campaign, five modules were tested: one module satisfied the requirement at 490~Gy but failed at 740~Gy, two modules satisfied the requirement at 720~Gy and failed at 780~Gy, and the remaining two modules satisfied the requirement at 780~Gy with no further irradiation performed. In the second campaign, six modules were tested: five modules satisfied the requirement at 600~Gy but failed at 700~Gy, while the remaining module satisfied the requirement at 700~Gy and failed at 800~Gy. Failures were observed in the output light intensity of the transmitters. Figure~\ref{SFP_plot_TID} shows the output light intensity for the second campaign, where the intensity dropped to 0~mW for five (six) modules after 700~Gy (800~Gy) irradiation.

Ten FTLX8574D3BCV modules were tested. All modules satisfied the requirement at 350~Gy, with failures observed between 400~Gy and 600~Gy. The failures occurred after a power cycle following irradiation, suggesting that the device configuration may have been affected.

Ten FSPP-H7-M85-X3DM modules were also tested. All modules satisfied the requirement at 200~Gy. Seven modules failed between 250~Gy and 350~Gy, while the remaining three modules were not irradiated beyond 200~Gy. For these modules, power cycling did not affect functionality. Communication tests performed with unirradiated counterpart SFP+ modules confirmed that the failures of the FSPP-H7-M85-X3DM were attributable to malfunctions in the receiver.

SFP+ optical transceivers are composite devices comprising multiple functional components, including a laser diode, a photodetector, memory, and driver/receiver circuitry. Radiation-induced failures can therefore manifest in different functional blocks depending on the module design and internal architecture.

\subsection{Clock jitter cleaner}

The devices irradiated in this study were the Si5344 and Si5395 clock jitter cleaners. Communication tests were conducted at 8.0~Gbps using the KC705 evaluation kit, with the Si5344 and Si5395 serving as the reference clock sources. The requirement was that the bit error rate remains negligible when transferring a 31-bit pseudorandom binary sequence.

A total of five Si5344 and two Si5395 devices were tested. For the Si5344, two devices satisfied the requirement at 490~Gy but could not be evaluated at 740~Gy due to a malfunction of another device mounted on the same evaluation board. One device was irradiated up to 480~Gy and satisfied the requirement, while the remaining two devices were irradiated up to 240~Gy and also satisfied the requirement.
For the Si5395, one device satisfied the requirement at 200~Gy but could not be evaluated at 300~Gy due to a malfunction of another device on the same board. The other device satisfied the requirement at 300~Gy but could not be evaluated at 400~Gy for the same reason.

\if0

\begin{table}[htbp]
\begin{center}
\caption{Result of TID test for jitter cleaner.
``Pass'' indicates that the chip is fine.
``--'' indicates that no measurement was performed
or the chip could not be evaluated due to a failure
for another element on the same board.
}
\scalebox{0.85}{
%\begin{tabular}{l|cccccccccccccc}\hline
%TID [Gy] & 200 & 240 & 300 & 400 & 480 & 490 & 740 \\ \hline \hline
%Si5344, chips 1, 2 & Pass & Pass & -- & -- & -- & Pass & ? \\ \hline
%Si5344, chip 3 & Pass & Pass & Pass & -- & Pass & -- & -- \\ \hline
%Si5344, chips 4, 5 & Pass & Pass & -- & -- & -- & -- & -- \\ \hline
%%Si5395, chip 0 & Pass & Pass & ? & -- & -- & -- & -- & -- & -- & -- \\ \hline
%Si5395, chip 1 & Pass & -- & ? & -- & -- & -- & -- \\ \hline
%Si5395, chip 2 & Pass & -- & Pass & ? & -- & -- & -- \\ \hline
%\end{tabular}
\begin{tabular}{l|cccccccccccc}\hline
TID [Gy] & 200 & 240 & 300 & 480 & 490 \\ \hline \hline
Si5344, chips 1, 2 & Pass & Pass & -- & -- & Pass \\ \hline
Si5344, chip 3 & Pass & Pass & Pass & Pass & -- \\ \hline
Si5344, chips 4, 5 & Pass & Pass & -- & -- & -- \\ \hline
%Si5395, chip 0 & Pass & Pass & ? & -- & -- & -- & -- & -- & -- & -- \\ \hline
Si5395, chip 1 & Pass & -- & -- & -- & -- \\ \hline
Si5395, chip 2 & Pass & -- & Pass & -- & -- \\ \hline
\end{tabular}
}
\label{Si_TID_Table}
\end{center}
\end{table}

\fi

\subsection{Optical fiber}

The optical fiber model irradiated in this study was 1-LC.P-LC.P-GI(PE-A10G)-DF-N-500. Two types of fiber cores were tested, each with a length of 500~m. The performance of the fibers was evaluated using a loopback test with the KC705 FPGA evaluation kit, and the light intensity after transmission through the fibers was also measured. Light intensity measurements were performed both during and after irradiation, with the light source located outside the irradiation area at all times.

\begin{figure}[htbp]
\begin{center}
\includegraphics[keepaspectratio, scale=0.30]{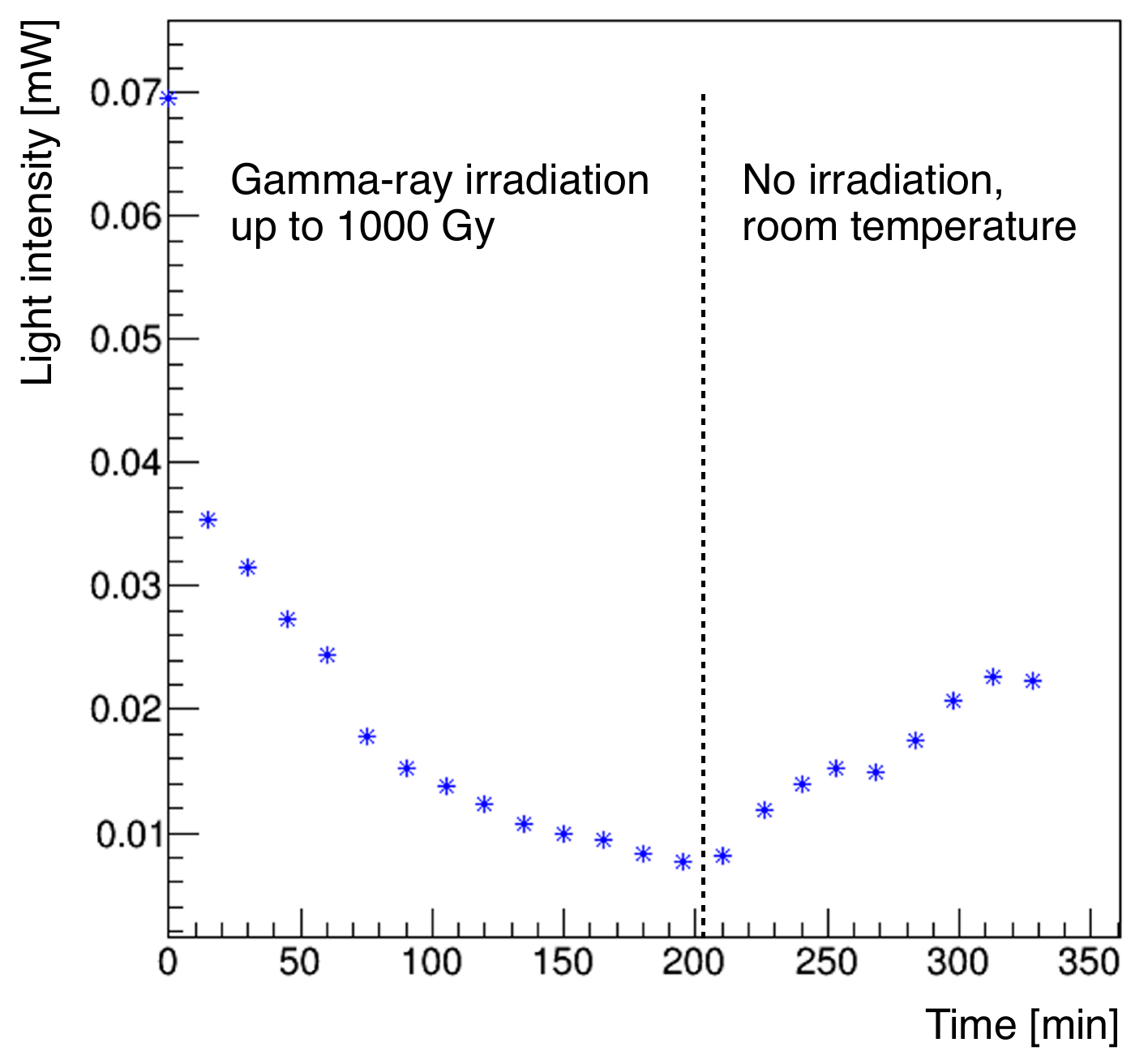}
\caption{Light intensity after transmission through the fiber as a function of time. Gamma-ray irradiation was performed for approximately 200~min, corresponding to a dose of up to 1000~Gy. After irradiation, the fiber was kept at room temperature with no further exposure.}
\label{Fibre_plot_TID}
\end{center}
\end{figure}

Figure~\ref{Fibre_plot_TID} shows the light intensity as a function of time for a tested fiber. A decrease in light intensity due to gamma-ray irradiation was observed, followed by partial recovery at room temperature when irradiation was stopped. One fiber was irradiated up to 2700~Gy, and the other up to 1200~Gy. Although a loss in light intensity was observed, it remained within the acceptable range for SFP+ modules, and communication tests showed no change in the bit error rate. These results confirm that the 1-LC.P-LC.P-GI(PE-A10G)-DF-N-500 optical fiber exhibits radiation tolerance at the level of O(1000)~Gy.

\subsection{Voltage reference}

\begin{figure}[htbp]
     \begin{center}
     \includegraphics[keepaspectratio, scale=0.35]{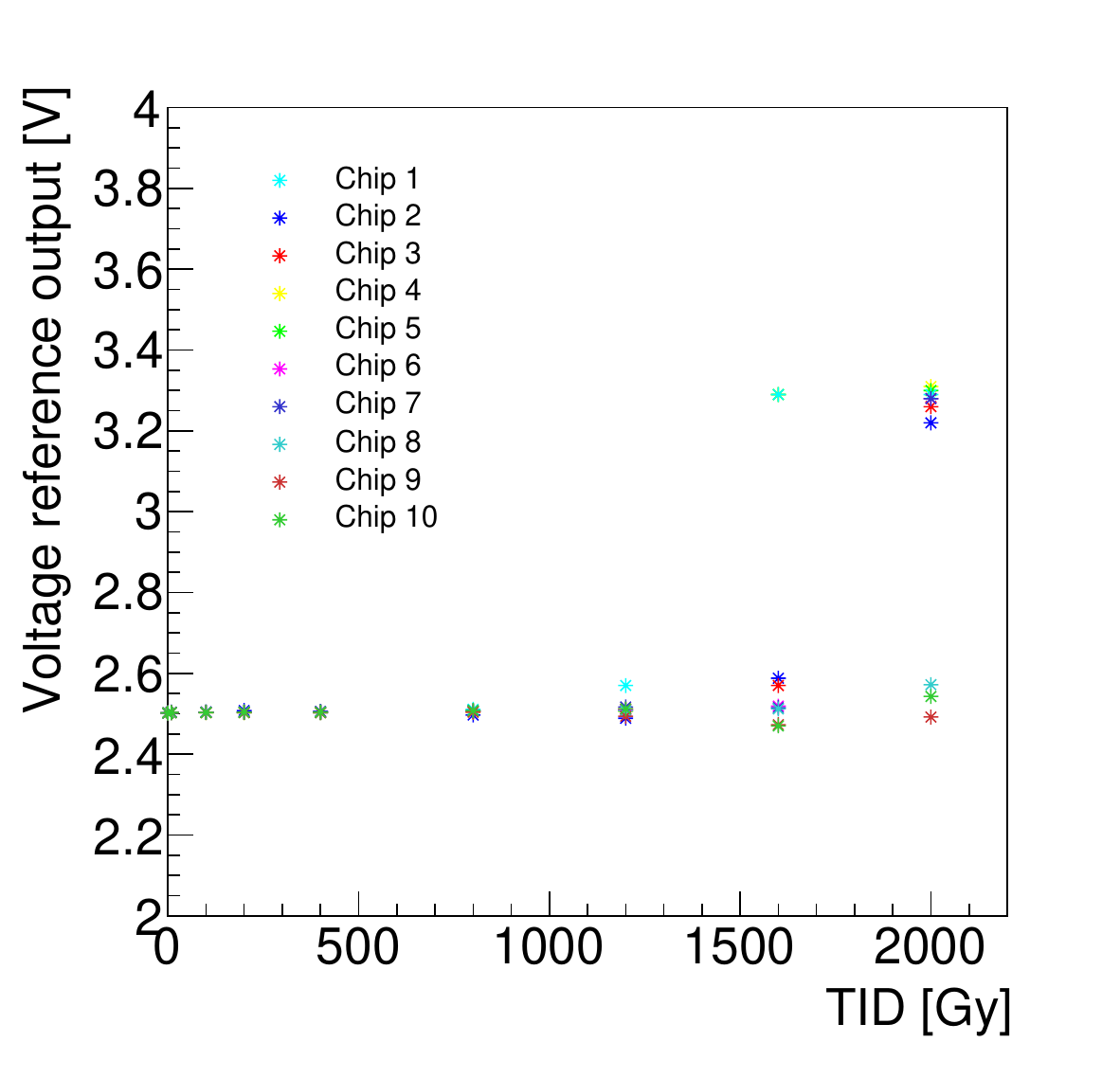}
     \caption{Output voltage  of REF2025 devices (nominal value: 2.5~V) as a function of the accumulated dose. Colors indicate individual chip identifiers.}
     \label{VERF_plot_TID}
     \end{center}
\end{figure}

The devices irradiated in this study were the REF2025, REF 5040, and REF5025 voltage reference models. We evaluated the change in output voltage before and after irradiation, and required that the relative change remain below 1\%.
For the REF2025, ten devices were tested. Figure~\ref{VERF_plot_TID} shows the measured output voltages. All tested REF2025 devices satisfied the requirement up to 800~Gy. One device failed at 1200~Gy, and all devices failed at 2400~Gy.
For the REF5040 and REF5025, ten devices of each model were tested. All tested devices met the requirement at 240~Gy, while failures were observed in the range of 270~Gy to 680~Gy.

\subsection{Operational amplifier}

The operational amplifier model irradiated in this study was the LM7322MM/NOPB. Ten devices were selected from each of two production lots. The relative change in output voltage was evaluated using an evaluation board and was required to be negligible. All tested devices satisfied this requirement at 1000~Gy. Additionally, five devices from each production lot were irradiated up to 2600~Gy, and all continued to meet the requirement.

\if0

\begin{table}[htbp]
\begin{center}
\caption{Result of TID test for LM7322MM/NOPB.
Ten chips were taken from each lot and irradiated.
``Pass'' indicates that the output voltages from the evaluation board BOB-14874,
where the chips originally on the boards were replaced by the chips from our lots,
are within the ranges of the specification and ``Fail'' indicates that they do not.
``--'' indicates that no measurement was performed at the corresponding dose.}
\begin{tabular}{l|cccccccccc}\hline
TID [Gy] & 11 & 1000 & 2600  \\ \hline \hline
Lot A, chips 1--5 & Pass & Pass & -- \\ \hline
Lot A, chips 6--10 & Pass & -- & Pass \\ \hline
Lot B, chips 1--5 & Pass & -- & Pass \\ \hline
Lot B, chips 6--10 & Pass & Pass & -- \\ \hline
\end{tabular}
\label{OpAmp_TID_Table}
\end{center}
\end{table}

\fi

\subsection{ADC}

The device irradiated in this study was ADS7953. The output values were compared with the expected values, with the requirement that the deviation of the output voltage be within 10 counts of the expected value. A total of ten devices were tested. One device satisfied the requirement at 250~Gy but could not be evaluated at 280~Gy due to a malfunction of the EEPROM on the same evaluation board. Another device satisfied the requirement at 180~Gy but failed at 270~Gy. The remaining eight devices satisfied the requirement at 240~Gy and were not subjected to further irradiation.

\if0

\begin{table}[htbp]
\begin{center}
\caption{Result of TID test for ADS7953.
``Pass'' indicates that the output is within 10 from the expected value,
where the output can take a value in the range from 0 to 4095.
``Fail'' indicates that the output is outside the range of $\pm 10$ from the expected value.
%``?'' indicates that EEPROM on the same board had a problem
%and ADC could not be tested.
``--'' indicates that no measurement was performed at the corresponding dose.}
\begin{tabular}{l|cccccccccccc}\hline
TID [Gy] & 180 & 210 & 240 & 250 & 270 \\ \hline \hline
Chip 1 & Pass & Pass & -- & Pass & -- \\ \hline
Chips 2--9 & Pass & -- & Pass & -- & -- \\ \hline
Chip 10 & Pass & -- & -- & -- & Fail \\ \hline
\end{tabular}
%\begin{tabular}{l|ccccccccccccc}\hline
%TID [Gy] & 180 & 210 & 240 & 250 & 270 & 280 \\ \hline \hline
%Chip 1 & Pass & Pass & -- & Pass & -- & ? \\ \hline
%Chips 2--9 & Pass & -- & Pass & -- & -- & -- \\ \hline
%Chip 10 & Pass & -- & -- & -- & Fail & -- \\ \hline
%\end{tabular}
\label{ADC_TID_Table}
\end{center}
\end{table}

\fi

\subsection{DAC}

The device irradiated in this study was DAC7678. The output voltage was compared with the programmed voltage, with the requirement that the deviation of the output voltage be within 0.01~V of the programmed voltage. A total of ten devices were tested. One device satisfied the requirement at 180~Gy but failed at 210~Gy. Another device was irradiated up to 270~Gy and satisfied the requirement. The remaining eight devices were irradiated up to 240~Gy, and all satisfied the requirement.

\if0

\begin{table}[htbp]
\begin{center}
\caption{Result of TID test for DAC7678.
``Pass'' indicates that the output voltage is within 0.01~V from the setup voltage,
and ``Fail'' indicates that the output voltage is outside the 0.01~V window.
``--'' indicates that no measurement was performed at the corresponding dose.}
\begin{tabular}{l|ccccccccccccc}\hline
TID [Gy] & 0-180 & 210 & 240 & 270 \\ \hline \hline
Chip 1 & Pass & Fail & -- & -- \\ \hline
Chips 2--9 & Pass & -- & Pass & -- \\ \hline
Chip 10 & Pass & -- & -- & Pass \\ \hline
\end{tabular}
\label{DAC_TID_Table}
\end{center}
\end{table}

\fi

\subsection{SD card} 

The SD card model irradiated in this study was the SDSD AF3-008G-I. Three functions were evaluated:
\begin{itemize}
\item Inserting the SD card into a laptop and verifying that it is recognized,
\item Checking that the hash value remains unchanged after irradiation,
\item Erasing all data, writing dummy data, and verifying the written data in the memory.
\end{itemize}
All functions were required to operate correctly without errors. Ten SD cards were tested, and all satisfied this requirement at 400~Gy. Failures were observed in the dose range between 600~Gy and 800~Gy.

\if0

\begin{table}[htbp]
\begin{center}
\caption{Result of TID test for SD card SDSDAF3-008G-I,
the type to be used for JATHub.
``Pass'' indicates that the SD card can be connected to the laptop,
the data stored before the irradiation were kept after the irradiation,
and the erase, write, and read operations work fine.
``Fail'' indicates that at least one of them fails.
``--'' indicates that no measurement was performed at the corresponding dose.}
\begin{tabular}{l|ccccccc}\hline
TID [Gy] & 400 & 600 & 800  \\ \hline \hline
Cards 1--5 & Pass & -- & Fail  \\ \hline
Card 6 & Pass & Pass & --  \\ \hline
Cards 7--9 & Pass & Fail & --  \\ \hline
Card 10 & Pass & Pass & --  \\ \hline
\end{tabular}
\label{SD_TID_Table}
\end{center}
\end{table}

\fi

\subsection{Flash memory}

The device irradiated in this study was the MX25L12845G M2I-08G flash memory. The functions evaluated after irradiation included reset, erase, write, and verification. All functions were required to operate correctly without errors. Eleven devices were tested, and all satisfied the requirement at 100~Gy. All tested devices failed at 150~Gy.

\if0

The model we irradiated is MX25L12845GM2I-08G. We tested reset, erase, write, and verify after irradiation.
%\begin{itemize}
%    \item reset 
%    \item erase, write, and verify 
%    \item irradiate
%\end{itemize}
``Pass" in the Table~\ref{QSPI_TID_Table} indicates all operations worked fine, and ``Fail" does not. It was confirmed that MX25L12845GM2I-08G has a radiation tolerance of up to at least 100~Gy.

\begin{table}[htbp]
\begin{center}
\caption{Result of TID test for flash memory MX25L12845GM2I-08G.
``Pass'' indicates that the erase, write, and verify operations work
and ``Fail'' indicates that at least one of them does not.
``--'' indicates that no measurement was performed at the corresponding dose.}
\begin{tabular}{l|cccccccccccccc}\hline
TID [Gy] & 100 & 125 & 135 & 140 & 145 & 150 \\ \hline \hline
Chips 1--6 & Pass & -- & -- & -- & -- & Fail \\ \hline
Chips 7, 8 & Pass & Pass & -- & -- & -- & Fail \\ \hline
Chip 9 & Pass & Pass & Pass & Pass & Fail & -- \\ \hline
Chips 10, 11 & Pass & Pass & Fail & -- & -- & -- \\ \hline
\end{tabular}
\label{QSPI_TID_Table}
\end{center}
\end{table}

\fi

\subsection{Low-dropout regulator}

The device irradiated in this study was the TPS7A85 low-dropout regulator. The output voltage and current were evaluated before and after irradiation. In Tables~\ref{LDO_TID_Table1} and \ref{LDO_TID_Table2}, ``Pass'' indicates that the output voltage and current remained within specifications for the conversions 3.3~V $\rightarrow$ 1.8~V (1.0~A output), 3.3~V $\rightarrow$ 1.0~V (0.55~A output), and 1.8~V $\rightarrow$ 1.2~V (0.66~A output), and that the ENABLE pin functioned correctly (ENABLE-DISABLE controllable).
``Pass$^{*}$'' indicates that the output voltage and current remained within specifications, but the ENABLE pin did not function (always ENABLE). Note that the ENABLE pin is unused in the TGC electronics.

\begin{table}[htbp]
\begin{center}
\caption{Results of the TID test for the TPS7A85 low-dropout regulator.
``Pass'' indicates that the output voltage and current remained within specifications and that the ENABLE pin functioned correctly.
``Pass$^{*}$'' indicates that the output voltage and current remained within specifications, but the ENABLE pin did not function.
``Fail'' indicates that the output voltage and/or current deviated from the expected values.
``--'' indicates that no measurement was performed at the corresponding dose.}
\begin{tabular}{l|cccccccccccc}\hline
TID [Gy] & 180 & 240 & 280 & 450 & 950 & 980 \\ \hline \hline
Chip 1 & Pass & -- & Pass & -- & -- & Pass$^{*}$ \\ \hline
Chips 2--4 & Pass & Pass$^{*}$ & -- & -- & -- & -- \\ \hline
Chips 5--8 & Pass & Pass & -- & -- & -- & -- \\ \hline
Chip 9 & Pass & -- & -- & Pass$^{*}$ & Pass$^{*}$ & -- \\ \hline
Chip 10 & Pass & -- & -- & Pass$^{*}$ & Fail & -- \\ \hline
\end{tabular}
\label{LDO_TID_Table1}
\end{center}
\end{table}

\begin{table}[htbp]
\begin{center}
\caption{Result of TID test for TPS7A85 (continued).}
\begin{tabular}{l|cccccccccc}\hline
TID [Gy] & 1040 & 1450 & 1950 \\ \hline \hline
Chip 1 & -- & -- & -- \\ \hline
Chips 2--4 & Pass$^{*}$ & -- & -- \\ \hline
Chips 5--8 & -- & -- & -- \\ \hline
Chip 9 & -- & Pass$^{*}$ & Fail \\ \hline
Chip 10 & -- & Fail & Fail \\ \hline
\end{tabular}
\label{LDO_TID_Table2}
\end{center}
\end{table}

\subsection{Low-dropout regulator for DDR}

The device irradiated in this study was the TPS51200 low-dropout regulator. The output voltage and current were evaluated before and after irradiation, with the requirement that the VTT and VTTREF output voltages remain within 5\% of their nominal values. A total of seven LDO devices were irradiated in two test campaigns. In the first campaign, two devices were tested: one satisfied the requirement at 960~Gy but failed at 1080~Gy, while the other satisfied the requirement at 240~Gy and was not subjected to further irradiation. In the second campaign, five devices were tested: four satisfied the requirement at 600~Gy but failed at 800~Gy, and the remaining device satisfied the requirement at 800~Gy and failed at 900~Gy.

\if0

\begin{table}[htbp]
\begin{center}
\caption{Result of TID test for the low-dropout regulator for DDR TPS51200 in the first campaign.
``Pass'' indicates that the VTT and VTTREF output voltages are within 5\% from the nominal values
and ``Fail'' indicates that they fail.
``--'' indicates that no measurement was performed at the corresponding dose.}
\begin{tabular}{l|cccccccccc}\hline
TID [Gy] & 240 & 960 & 1080  \\ \hline \hline
Chip 1 & Pass& Pass & Fail \\ \hline
Chip 2 & Pass & -- & -- \\ \hline 
\end{tabular}
\label{DDRLDO_TID_Table1}
\end{center}
\end{table}

\begin{table}[htbp]
\begin{center}
\caption{Result of TID test for the low-dropout regulator for DDR TPS51200 in the second campaign.
``Pass'' indicates that the VTT and VTTREF output voltages are within 5\% from the nominal values
and ``Fail'' indicates that they fail.
``--'' indicates that no measurement was performed at the corresponding dose.}
\begin{tabular}{l|cccccccc}\hline
TID [Gy] & 600 & 800 & 900  \\ \hline \hline
Chips 1, 2 & Pass & Fail & Fail \\ \hline
Chip 3 & Pass & Pass & Fail \\ \hline
Chips 4, 5 & Pass & Fail & -- \\ \hline
\end{tabular}
\label{DDRLDO_TID_Table2}
\end{center}
\end{table}

\fi

\subsection{Summary}

The full results of the TID tests are summarized in Table~\ref{TID_Sumamry_Table}. The ``confirmed'' values indicate the maximum dose at which functionality was verified for the most sensitive elements among the tested devices. From these results, we conclude that all tested devices exhibit sufficient TID tolerance for use in the TGC electronics at the HL-LHC.

\begin{table*}[htbp]
\centering
\caption{Summary of TID test results.
``Confirmed'' indicates the maximum dose at which proper functionality was verified for the worst-performing device among the tested samples.
A lot safety factor of ${\rm SF}_{\rm lot} = 1$ is applied and a relaxed requirement is adopted for REF2025, LM7322MM/NOPB, and MX25L12845GM2I-08G, as the production reel used for assembly is identical to that used in the TID test.
The requirements for SDSDAF3-008G-I and TPS51200 are also relaxed, since these devices are used exclusively on the JATHub boards.}
\begin{adjustbox}{width=\textwidth,keepaspectratio}
\begin{tabular}{l|l|c|c|c}
\hline
Type                 & Model                            & Number of tested & Requirement [Gy] & Confirmed [Gy]\\ \hline \hline
SFP+ transceiver     & AFBR-709SMZ                & 11            & 33          & 490 \\ \hline
SFP+ transceiver     & FTLX8574D3BCV            & 10               & 33          & 350       \\ \hline
%SFP+ transceiver     & Coherent FTLX8573D3BTL            & 1                & 33          & 240       \\ \hline
SFP+ transceiver     & FSPP-H7-M85-X3D                & 10            & 33          & 200 \\ \hline
%SFP+ transceiver     & FS SFP-10GSR-85                & 17            & 11          & 83 \\ \hline
Clock jitter cleaner & Si5344                         & 5                & 33          & 240       \\ \hline
Clock jitter cleaner & Si5395                           & 2                & 33          & 200       \\ \hline
Optical fiber        & 1-LC.P-LC.P-GI(PE-A10G)-DF-N-500 & 2                & 33          & 1200       \\ \hline
Voltage reference    & REF2025                          & 10               & 11           & 800      \\ \hline
Voltage reference     & REF5040                          & 10               & 33          & 240       \\ \hline
Voltage reference    & REF5025                          & 10               & 33          & 240       \\ \hline
Operational amplifier                & LM7322MM/NOPB                    & 20               & 11          & 1000      \\ \hline
ADC                  & ADS7953                          & 10               & 33          & 180       \\ \hline
DAC                  & DAC7678                          & 10               & 33          & 180       \\ \hline
SD card              & SDSDAF3-008G-I                   & 10               & 18          & 400       \\ \hline
Flash memory    & MX25L12845GM2I-08G               & 11               & 11          & 100       \\ \hline
Regulator         & TPS7A85                          & 10               & 33          & 240       \\ \hline
Regulator for DDR & TPS51200                         & 7            & 18          & 240 \\ \hline
\end{tabular}
\end{adjustbox}
\label{TID_Sumamry_Table}
\end{table*}

\section{Result of the NIEL tests}

The full results of the neutron irradiation tests are summarized in Table~\ref{NIEL_Summary_Table}. The term ``confirmed'' indicates the maximum irradiation level at which the devices remained fully functional. The functionality of the tested components was evaluated using the same procedures as those employed for the TID tests.
%For the SFP+ transceiver FTLX8573D3BTL, the eye pattern of the communication test was intact at $11 \times 10^{12}$~n$_{\rm1MeV}$~cm$^{-2}$, but degradation was observed at $13 \times 10^{12}$~n$_{\rm1MeV}$~cm$^{-2}$.
No failures were observed for the tested devices.
From these results, we conclude that all tested devices exhibit sufficient NIEL tolerance for application in the TGC frontend electronics at the HL-LHC.

\begin{table*}[thbp]
\centering
\caption{Summary of the NIEL test results.
%For the SFP+ transceiver FTLX8573D3BTL, the eye pattern remained intact at $11 \times 10^{12}$~n$_{\rm1MeV}$~cm$^{-2}$, but degradation was observed at $13 \times 10^{12}$~n$_{\rm1MeV}$~cm$^{-2}$.
No failures were observed for the tested devices. The column labeled ``requirement'' indicates the NIEL tolerance required, including safety factors. The column labeled ``confirmed'' shows the 1-MeV neutron equivalent fluence applied in this test, with ranges provided for different device types.}
\begin{adjustbox}{width=\textwidth,keepaspectratio}
\begin{tabular}{l|l|c|c|c} \hline 
Type & Model & Number of tested & Requirement & Confirmed \\
 & & & [$\times 10^{12}$~n$_{\rm1MeV}$ $\rm{cm}^{-2}$] & [$\times 10^{12}$~n$_{\rm1MeV}$ $\rm{cm}^{-2}$] \\ \hline \hline
SFP+ transceiver & AFBR-709SMZ & 2 & 1.3 & 4.6--12 \\ \hline
SFP+ transceiver & FTLX8574D3BCV & 1 & 1.3 & 4.6 \\ \hline
%SFP+ transceiver & FTLX8573D3BTL & 1 & 1.3 & 11 \\ \hline
SFP+ transceiver & FSPP-H7-M85-X3D & 10 & 1.3 & 0.8--11 \\ \hline
%SFP+ transceiver & SFP-10GSR-85 & 4 & 0.4 & 6--20 \\ \hline
Voltage reference & REF2025 & 10 & 0.4 & 1.1--2.2 \\ \hline
Operational amplifier & LM7322MM/NOPB & 10 & 0.4 & 1.3--3.0 \\ \hline
DAC & DAC7678 & 1 & 1.3 & 5.3 \\ \hline
Flash memory & MX25L12845GM2I-08G & 9 & 0.4 & 3.6--11 \\ \hline
Regulator & TPS7A85 & 1 & 1.3 & 5.3 \\ \hline
Regulator for DDR & TPS51200 & 3 & 0.7 & 1.9--3.9 \\ \hline
\end{tabular}
\end{adjustbox}
\label{NIEL_Summary_Table}
\end{table*}

\section{Conclusion}
We have performed gamma-ray and neutron irradiation tests on COTS components intended for the TGC frontend electronics of the ATLAS experiment at the HL-LHC. Gamma-ray irradiation was carried out at the Cobalt-60 facility of Nagoya University, while neutron irradiation was performed at the Tandem Accelerator of Kobe University. 
All tested components met the radiation tolerance requirements for ten years of HL-LHC operation, including the applied safety factors, demonstrating sufficient tolerance to both TID and NIEL.
For the gamma-ray tests, we determined the maximum dose at which the components remained fully functional. In the neutron irradiation tests, all components continued to operate correctly after exposure up to O($10^{12}$) n$_{\rm1MeV}$ $\rm{cm}^{-2}$.

The full production of the TGC frontend electronics was completed using Si5395, REF2025, LM7322MM/NOPB, ADS 7953, DAC7678, SDSDAF3-008G-I, MX25L12845GM2I-08G, TPS7A85, and TPS51200.
The final quality verification, including the TID test performed on a fully assembled board, was successfully completed in 2025.
The SFP+ optical transceiver, which is insertable to the boards, follows a separate procurement timeline.
The specific SFP+ model is selected based on an evaluation of Single Event Effects to further enhance the overall robustness of the system.

\section*{Acknowledgement}

This work was supported by JSPS KAKENHI Grant Numbers 16H06493, 21H05085, and 22H04944. We are grateful to J. Kumagai and S. Imai for their support in managing the Cobalt-60 facility at Nagoya University. We also express our sincere thanks to the staff of the Tandem Accelerator Laboratory at Kobe University.

\section*{Declaration of generative AI and AI-assisted technologies in the manuscript preparation process}

During the preparation of this work the authors used ChatGPT (OpenAI) in order to improve the clarity and readability of the manuscript. After using this tool/service, the authors reviewed and edited the content as needed and take full responsibility for the content of the published article.

%\label{}

%% The Appendices part is started with the command \appendix;
%% appendix sections are then done as normal sections
%% \appendix

%% \section{}
%% \label{}

%% If you have bibdatabase file and want bibtex to generate the
%% bibitems, please use
%%
%%  \bibliographystyle{elsarticle-num} 
%%  \bibliography{<your bibdatabase>}

\begin{thebibliography}{99}

\bibitem[1]{ATLAS}
ATLAS Collaboration, ``The ATLAS Experiment at the CERN Large Hadron Collider'', 2008 JINST 3 S08003.

\bibitem[2]{Higgs}
ATLAS Collaboration, ``Observation of a new particle in the search for the Standard Model Higgs boson with the ATLAS detector at the LHC'', Phys. Lett. B 716, 1--29 (2012).

\bibitem[3]{Coupling}
ATLAS Collaboration, ``A detailed map of Higgs boson interactions by the ATLAS experiment ten years after the discovery'', Nature 607, 52--59 (2022).

\bibitem[4]{HL-LHC}
I.~B.~Alonso, O.~Br\"{u}ning, P.~Fessia, M.~Lamont, L.~Rossi, L.~Tavian, and M.~Zerlauth, ``High-Luminosity Large Hadron Collider (HL-LHC): Technical design report'', CERN-2020-010.

\bibitem[5]{MUON_TDR}
ATLAS Collaboration, ``ATLAS muon spectrometer: Technical Design Report'', CERN-LHCC-97-022; ATLAS-TDR-10.

\bibitem[6]{PhaseII_TDR}
ATLAS Collaboration, ``Technical Design Report for the Phase-II Upgrade of the ATLAS Muon Spectrometer'', CERN-LHCC-2017-017; ATLAS-TDR-026.

\bibitem[7]{Geant4}
GEANT4 Collaboration, ``GEANT4 --- a simulation toolkit'', Nuclear Inst. and Methods in Physics Research, A 506, 250--303 (2003).

\bibitem[8]{TwoMeV}
T.~Inada and K.~Kawachi, ``Neutrons from Thick Target Beryllium $(d, n)$ Reactions at 1.0~MeV to 3.0~MeV'', Journal of Nuclear Science and Technology, 5, 22--29 (1968).

\bibitem[9]{NeutronFlux}
Y.~Nakazawa, Y.~Fujii, E.~Hamada, M.~Lee, Y.~Miyazaki, A.~Sato, K.~Ueno, H.~Yoshida, and J.~Zhang, ``Radiation study of FPGAs with neutron beam for COMET Phase-I'', Nuclear Inst. and Methods in Physics Research, A 936, 351--352 (2019).

\end{thebibliography}

%% else use the following coding to input the bibitems directly in the
%% TeX file.

%\begin{thebibliography}{00}

%% \bibitem{label}
%% Text of bibliographic item

%\bibitem{}

%\end{thebibliography}
\end{document}